\documentclass{aa}
\usepackage{epsf}
\usepackage{aafonts}

\def\etal{{et~al.}}

\def\ApJL{ApJ}

\def\ApSS{Astrophys$\,$Space$\,$Sci}

\def\lesssim{\mathrel{\hbox{\rlap{\hbox{\lower4pt\hbox{$\sim$}}}\hbox{$<$}}}}
\def\gtrsim{\mathrel{\hbox{\rlap{\hbox{\lower4pt\hbox{$\sim$}}}\hbox{$>$}}}}

\begin{document}

\thesaurus{13(13.07.1)}

\title {GRANAT/SIGMA Observation of the Early Afterglow from GRB~920723 in
  Soft Gamma-Rays}

\author {R.~A.~Burenin\inst{1,2}, A.~A.~Vikhlinin\inst{1}, 
M.~R.~Gilfanov\inst{1,3}, O.~V.~Terekhov\inst{1,2}, 
 A.~Yu.~Tkachenko\inst{1,2},
S.~Yu.~Sazonov\inst{1,2}, E.~M.~Churazov\inst{1,3},
 R.~A.~Sunyaev\inst{1,3}, 
P.~Goldoni\inst{4}, A.~Claret\inst{4}, A.~Goldwurm\inst{4}, 
J.~Paul\inst{4}
J.~P.~Roques\inst{5}, E.~Jourdain\inst{5}, F.~Pelaez\inst{6}, 
G.~Vedrenne\inst{5}  }

\institute{
Space Research Institute, Russian Academy of Sciences, Profsoyuznaya
84/32, 117810 Moscow, Russia 
\and
visiting Max-Planck-Institut f\"ur Astrophysik
\and
Max-Planck-Institut f\"ur Astrophysik,
Karl-Schwarzschild-Str. 1, 85740 Garching bei Munchen, Germany 
\and
CEA/DSM/DAPNIA/SAp, Centre d'Etudes de Saclay, 91191, Gif-sur-Yvette, Cedex,
France 
\and
Centre d'Etude  Spatiale des Rayonnements (CNRS/UPS) 9, avenue 
du Colonel Roche, BP 4346 31028 Toulouse Cedex, France
\and
Department of Physics and Astronomy, Mississippi State University, MS 39762
}


\maketitle

\begin{abstract}
  
  We present a GRANAT/SIGMA observation of the soft gamma-ray afterglow
  immediately after GRB~920723. The main burst is very bright. After
  $\sim6\,$s, the burst light curve makes a smooth transition into an
  afterglow where flux decays as $t^{-0.7}$. The power-law decay lasts for
  at least 1000$\,$s; beyond this time, the afterglow emission is lost in
  the background fluctuations. At least ${\sim}20\%$ of main burst energy is
  emitted in the afterglow. At approximately $\sim6\,$s after the trigger,
  we also observe an abrupt change in the burst spectrum. At $t<6\,$s, the
  ratio of 8--20 and 75--200~keV fluxes corresponds to the power law
  spectral index $\alpha=0.0-0.3$. At $t=6\,$s, the value of $\alpha$
  increases to $\alpha\approx 1$ and stays at this level afterwards. The
  observed afterglow characteristics are discussed in connection with the
  relativistic fireball model of gamma-ray bursts.


\end{abstract}

\markboth{R.A. Burenin et al.: GRB 920723 Soft Gamma-Ray Afterglow}{}

\section{Introduction}

Fast and accurate localizations of gamma-ray bursts by {\em Beppo}\-SAX helped
to establish the connection of GRB with the sources of decaying X-ray,
optical, and radio emission (e.g. Costa \etal\ 1997, Van Paradijs \etal\ 
1997, Frail \etal\ 1997). X-ray afterglows were found in 15 of 19
well-localized bursts; in most cases, the X-ray flux decayed as a power law
of time, $t^{-\beta}$, with $\beta$ ranging from $-1.1$ (GRB~970508, Piro
\etal\ 1998) to $-1.57$ (GRB~970402, Nicastro \etal\ 1998). The power law
decay of flux also is observed in the optical (e.g. Wijers \etal\ 1997,
Sokolov \etal\ 1998). This is a characteristic prediction of the
relativistic fireball model of GRB (M\'esz\'aros, Rees\ 1993, M\'esz\'aros\ 
1997, Waxman\ 1997, Sari \etal\ 1998). Indeed, the energy release in some
GRB is enormous and sufficient to power the relativistic fireball (Kulkarni
\etal\ 1998).
The fireball observations immediately after the burst, when the temperature
and density are at maximum, are of great interest. Unfortunately, it has
been impossible to observe the afterglows in the radio, optical, or X-rays
earlier than approximately 10 hours after the burst.

Some earlier observations indicated that the afterglows could immediately
follow some GRB. There were detections of X-ray emission lasting for tens of
seconds after the main burst was finished in gamma-rays (Sunyaev \etal\ 
1990, Murakami \etal\ 1991, Terekhov \etal\ 1993, Sazonov \etal\ 1998).  PVO
observatory observed a faint gamma-ray emission over $\sim1000$~s after the
long, $200\,$s, burst GRB~840304 (Klebesadel 1992). The presence of slowly
fading soft gamma-ray (100--500~keV) emission was found in about $\sim10\%$
of bursts detected by GRANAT/PHEBUS (Tkachenko \etal\ 1995). Hard gamma-ray
photons (0.2--10~GeV) were detected during $1.5$ hours after GRB~940217 by
EGRET telescope (Hurley \etal\ 1994).

We present here a detailed analysis of the GRB~920723 light curve, which
reveals a soft gamma-ray afterglow with flux decaying as a power law $\sim
t^{-0.7}$ during at least 1000~s after the main burst.

\begin{figure}[htb] 
  \vskip -37mm
  \centerline{\epsfxsize=1.\linewidth \epsffile{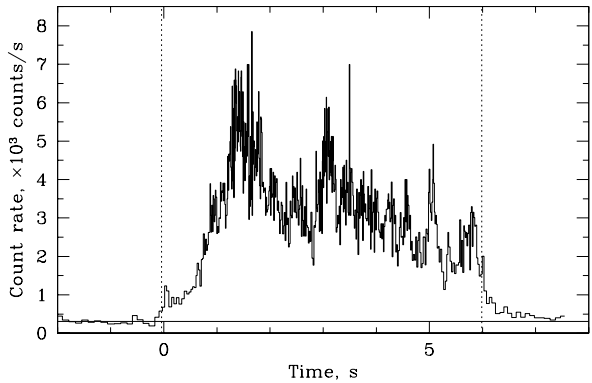}}
  \vskip -6mm
  \caption{ GRB~920723 light curve with time resolution $\lesssim0.1$ in
    35--300~keV band. The reference time is at the burst trigger. Vertical
    dotted lines represent the reference times of the light curve in
    logarithmic coordinates (Fig.~2 and 4). }
  \end{figure}

\section{Observations}

SIGMA is the coded-mask telescope with a $15'$ angular resolution operating
in the 35--1300~keV energy band (Paul \etal\ 1991). Typically, SIGMA
performs uninterrupted 20--30\,h observations, during which the telescope
pointing is maintained with a $30'$~accuracy (but known to within $15''$).
Although the telescope field of view is only $11.4^\circ\times10.5^\circ$
(FWHM), some fraction of gamma-rays from sources closer than $35^\circ$ to
the pointing direction reaches the detector through the gaps in the passive
shield and produces arc-shaped images. This ``secondary optics'' (or
``sidelobes'') was described in detail by Claret \etal\ (1994a) along with
the appropriate analysis techniques.

GRB~920723 was observed by SIGMA through the secondary optics. An
$\sim1^\circ$ localization was obtained from this observation (Claret \etal\ 
1994b). GRB~920723 is one of the brightest bursts observed by GRANAT
instruments and the brightest detected by SIGMA. The burst was triggered at
$20^{\rm h}03^{\rm m}08^{\rm s}.3$ UT and lasted for about 6~s.  The WATCH
all-sky monitor provided a $0.2^\circ$ localization (Sazonov 1998) and
observed the fading X-ray emission in the 8--20~keV band during more than
$40$~s after the main burst (Terekhov \etal\ 1993). PHEBUS measured the peak
burst flux $5\times10^{-5}$~erg~s$^{-1}$~cm$^{-2}$ and fluence
$1.4\times10^{-4}$~erg~cm$^{-2}$ in the 100--500~keV energy band (Terekhov
\etal\ 1995).

The SIGMA data allows the measurements of the burst light curve with better
than $0.1$~s time resolution (depending on flux) during $7.5$~s after the
trigger. In addition, the count rate in four wide energy bands (35--70,
70--150, 150--300, 300--600~keV) is recorded with the 4~s time resolution
over the entire observation. With these data, it is possibile to study the
burst emission long after the trigger. Below, we use only the first three
energy channels, because the last one is plagued by low sensitivity.  The
peak burst count rate in the 35--300~keV band was 7900~cnt~s$^{-1}$, much
higher than the average background rate 310~cnt~s$^{-1}$.

During the observation of July 23, 1992, SIGMA was pointed to Her~X-1. The
pulsar was in eclipse and was not detected. The $3\sigma$ upper limit on its
35--70~keV flux averaged over the observation, was $0.25$~cnt~s$^{-1}$. The
Her~X-1 spectrum is known to be very soft (the 20--100~keV photon index is
$-4.4$), and therefore its flux is negligible above 70~keV. The pulsar was
in eclipse between 12000~s before the burst and 9000~s after the burst.
Therefore, it could not cause any significant variability of the SIGMA count
rate during the reported observation. No other known bright sources were
visible through either primary or secondary optics of the SIGMA telescope.
GRANAT operates on the high apogee orbit and, during the observation, was
not influenced by the Earth radiation belts or other magnetospheric
anomalies (such as the South Atlantic Anomaly). As a result, the SIGMA
background usually does not show any significant variations on the time
scales shorter than $\sim10^3$~s.  Therefore, it can be accurately modeled
by a low degree polynomial.


\section{Results}

Usually, sources contribute only a small fraction of the total SIGMA count
rate. Therefore, the correct background subtraction is vital for the source
variability studies. We modeled the background using Chebyshev polynomials.
A complete description of the SIGMA background subtraction techniques is
presented elsewhere (Burenin et al.\ 1999); this analysis has shown that the
background variations around the subtracted value in excess of Poisson noise
are smaller than 0.6 cnt~s$^{-1}$ on the 300~s time scale (on the $95\%$
significance level).

Figure~1 shows the burst light curve in the 35--300~keV band. There is a
small peak in burst light curve just after the trigger. Within the first
second after the trigger, the burst flux rose rapidly. Over the next five
seconds, it remained at approximately the same level, showing a strong
variability at all resolved time scales. At approximately 6~s after the
trigger, the flux started to decline rapidly.

\begin{figure}[tb] 
\vskip -5mm
\centerline{\epsfxsize=\linewidth \epsffile{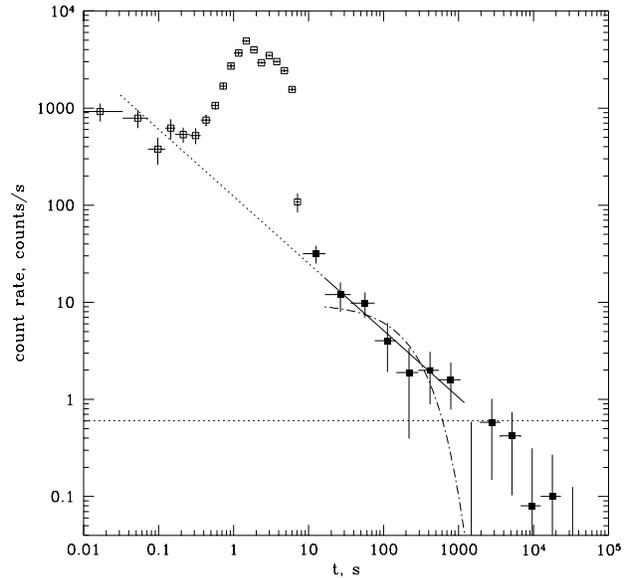}}
\vskip -5mm \caption{ The background-subtracted burst light curve. Zero
  time is at the burst trigger. The filled squares show the count rate
  measured with the 4~s resolution, and the open squares --- the data with
  good time resolution.  The horizontal dotted line represents a $95\%$
  upper limit on the possible internal background variations on the 300~s
  time scale (note that Poisson variations are already included in the
  error bars).}
\end{figure}

\begin{figure}[tb] 
  \vskip -32mm
  \centerline{\epsfxsize=\linewidth \epsffile{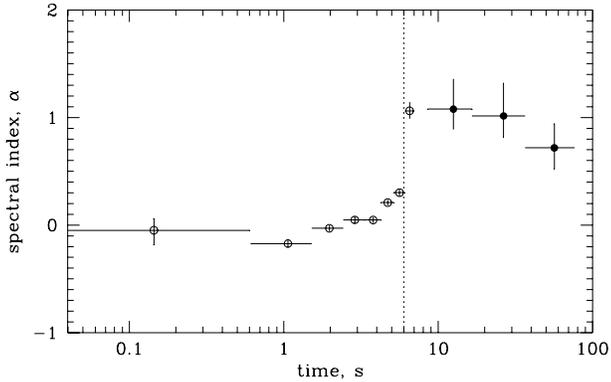}}
  \vskip -5mm
  \caption{Time history of the effective spectral index in the 8--200 keV
    energy band. Zero time is at the burst trigger. Vertical dotted line
    represents the moment $t=6\,$s from the trigger, when the burst flux
    began the gradual decline.}
\end{figure} 

\begin{figure}[tb] 
  \centerline{\epsfxsize=\linewidth \epsffile{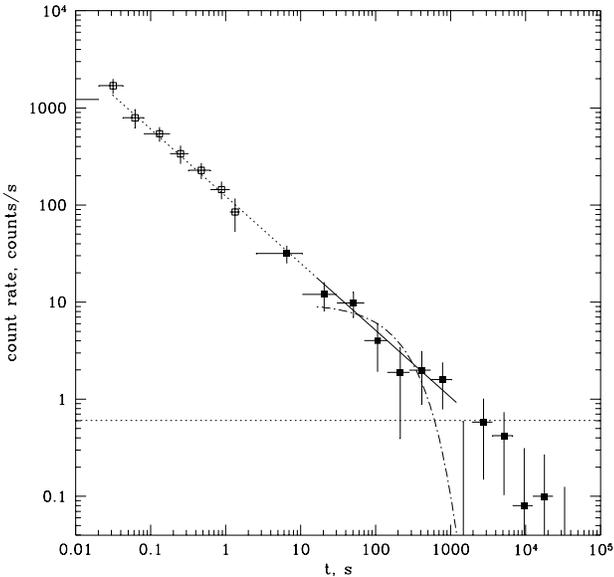}}
  \vskip -5pt
  \caption{Same as Fig.~2, but the reference time was set at
    6~s after the trigger. The main burst is not shown here because it is at
    $t<0$ with this choice of reference time.}
\end{figure} 
  
Figure 2 shows the burst light curve in logarithmic coordinates of both time
and flux. The shape of the light curve in these coordinates strongly depends
on the choice of the reference time. In Fig~2, the reference time is chosen
at the moment of the burst trigger. There appears to be a power-law decay of
flux starting at 10--20~s after the trigger.  This behavior is consistent
with GRB entering the stage of self-similar fireball expansion soon after
the main burst. It should be emphasized that the self-similar behavior is
expected only on time scales much larger than those of the main energy
release.  Therefore, we used the data in the $20$--$1000$~s time interval
for the power-law modeling of the light curve.

The solid line in Fig.~2 shows the power law fit in the time interval
$20$--$1000$~s; the dash-dotted line shows the exponential fit in the same
interval. The reduced $\chi^2$ is 1.5 and 6.5 (4 dof) for the power law and
exponential models, respectively. The power law adequately describes the
data and results in a better fit than the exponent. The best fit power law
index is $-0.69\pm0.17$ ($\Delta\chi^2=2.7$ for the index $-1$). This power
law tail contains at least $\sim20\%$ of the main burst fluence.  Applying
the same procedure to the three wide energy channels separately, we obtained
the power law indices $-1.29\pm0.55$, $-0.64\pm0.19$, and $-0.41\pm0.6$ in
the 35--70, 70--150, and 150--300~keV energy bands, respectively.
Interestingly, the extrapolation of the power law (shown by the dotted line
in the Fig.~3) points to the small peak near the beginning of the main burst
(Fig.~1).

The spectral evolution of the burst flux can be characterized by the ratio
of the 8--20~keV flux measured by WATCH (Terekhov et al.\ 1993, Sazonov et
al.  1998) and the SIGMA flux in the 75--200~keV band. Standard SIGMA
calibration does not apply to the secondary optics flux, so we used the
PHEBUS measurement of the main burst flux to find the conversion coefficient
between SIGMA counts and flux. Figure~3 shows the observed ratio of 8--20
and 75--200~keV fluxes expressed in terms of equivalent spectral index in
the 8--200~keV energy range (i.e., the spectral index of an
$F_\nu\propto\nu^{-\alpha}$ spectrum with the same hardness ratio as
observed). The time behavior of $\alpha$ indicates that the afterglow
spectrum is significantly softer than that of the main burst.  During the
main burst (in the 0--6~s time interval), the usual hard-to-soft evolution
of the GRB spectrum is observed (e.g. Ford \etal\ 1995). Near the start of
the gradual decline of the burst flux, 6~s after the trigger (shown by the
vertical dotted line in Fig.~1 and 3), $\alpha$ changes abruptly from
$\approx0.3$ to $\approx 1$.

It is interesting to examine the light curve with the reference time chosen
at $t=6\,$s after the trigger because this moment can be singled out in both
flux and spectral history of the burst; the result is presented in Fig.~4.
With this choice of zero time, the data in the 0.01--20~s time interval lies
on the extrapolation of the power law fit from the 20--1000~s time interval.
Adding these data to the fit results in the power law index $-0.70\pm0.03$.
Note that in this case, the power law flux decay is observed over
approximately four orders of magnitude of time.

\section{Discussion}

We presented a high sensitivity observation of the GRB~920723 light curve in
the soft gamma-ray band. The stable background of SIGMA allows detection of
the burst emission on the level of better than 1/1000 of the peak intensity.
A similar analysis would be complicated with BATSE because the background is
less stable and because the source is eclipsed by Earth every several
thousands seconds. We were able to detect the burst afterglow extending up
to $\sim1000$~s after the main burst. There is a continuous transition of
the main burst to its power law afterglow (Fig.~1 and 2). The afterglow
spectrum is significantly softer than that of the main burst. An abrupt
change in the burst spectrum occurs at approximately the same moment when
the power law decay of flux seems to start, at $t\approx 6\,$s after the
trigger. 

The behavior of GRB afterglows in the lower energy bands and at $t\gtrsim
3\times 10^4$~s can be explained by the synchrotron emission of electrons
accelerated in external shocks generated by relativistically expanding
fireball colliding with the interstellar medium (e.g. Wijers \etal\ 1997,
Waxman\ 1997, Sari \etal\ 1998, Wijers and Galama\ 1998). In the framework
of this model, the spectral flux at the observed frequency $\nu$ is given by
$F_\nu\propto\nu^{-\alpha}t^{-\beta}$, where $\alpha$ and $\beta$ are
constant and depend only on the spectral index of electrons on sufficiently
late stages of the fireball evolution. For GRB~920723 we obtain
$\alpha=1\pm0.2$ and $\beta=0.69\pm0.17$. Both the spectrum and the light
curve of GRB~920723 seem to be considerably flatter than that of X-ray
afterglows observed for other gamma-ray bursts at $t>3\times 10^{4}$ s ---
$\alpha=1.4$--$1.7$ and $\beta=1.1$--$1.6$ (e.g., in't Zand \etal\ 1998,
Piro \etal\ 1998, Nicastro \etal\ 1998). Furthermore, the flux decay
observed by SIGMA is flatter than $t^{-1}$ (at $\sim90$\% confidence), i.e.\ 
the total flux diverges if extrapolated to $t \rightarrow \infty$. 
This suggests that the afterglow light curve should steepen
at some moment during or after the SIGMA observation.

In the relativistic fireball model, the afterglow light curve and the energy
spectrum should steepen simultaneously at the moment $t_m$ when the maximum
in the electron spectrum, $E_m$, passes through the SIGMA bandpass (we
assume below that $t_m$ corresponds to $E_m=100\,$keV). At later stages of
the fireball evolution, indices $\alpha$ and $\beta$ do not change with
time. Since the light curve steepening after 100--1000$\,$s is required by
our data at $\sim 90\%$ confidence, it can be suggested that
$t_m>100-1000\,$s. Also, if indeed $t_m$ is $>\!100\,$s, our flat spectrum
may become consistent with the parameters of X-ray afterglows (see above)
because the spectrum softens after $t_m$. In the adiabatic fireball, $t_m$
can be estimated as $t_m\approx 140\, \varepsilon_B^{1/3}
\varepsilon_e^{4/3} E_{53}^{1/3}$~s, where $E_{53}\times10^{53}\,$ergs is
the total energy release, and $\varepsilon_e<1$ and $\varepsilon_B<1$ are
the fractions of the electron and magnetic field energy in the total shock
energy, respectively (Sari et al.\ 1998). The value of $t_m$ not much less
than $100\,$s would not be strongly inconsistent with the SIGMA data.
However, using $\varepsilon_e$ and $\varepsilon_B$ estimates from the
parameters of radio, optical, and X-ray afterglows of GRB~970508 and
GRB~971214 at $t\sim10^5-10^6\,$s (Wijers \& Galama\ 1998), we obtain
$t_m\sim 3$~s, which does seem to contradict our data at $\sim 90\%$
confidence level. This may indicate a large diversity of the fireball
parameters in different bursts or some problems of a simple model of a
spherically symmetric fireball in explaining the early stages of the
gamma-ray burst afterglows.

SIGMA data provides the first convincing observation of the power law
afterglow in the soft gamma-rays and immediately after the burst. A very
important issue is whether such afterglows are common. A preliminary
analysis of other SIGMA bursts revealed no other convincing afterglows,
primarily because of the faintness of other bursts; on the basis of SIGMA
data alone, we cannot rule out that the soft gamma-ray afterglow is a common
phenomenon.  A preliminary analysis of the PHEBUS data confirms the
detection of the afterglow in GRB~920723 and reveals a similar afterglow in
GRB~910402 (Tkachenko \etal\ 1998). The results of our systematic search for
soft gamma-ray afterglows in the GRANAT data will be presented in the
future.

\section{Acknowledgments}
This work was supported by RBRF grants 96--02--18458 and 96--15--96930.

\appendix
\section{Background modeling}

{\small (This section does not appear in the Journal version)\par}
\bigskip

We start with fitting the background by the Chebyshev polynomial. We exclude
the data between 1000~s before the burst and 4000~s after the burst. We
increase the order of the polynomial until the F-test indicates that no
additional powers of $t$ are necessary; we find that the polynomials of the
second and third orders are required to describe the background in spectral
channels 1\&2 and 3, respectively.

Since the fit is made for the entire observation, the uncertainty of the fit
value, arising from statistical uncertainties in the polynomial
coefficients, is negligible in any small part of the observation --- at
least compared to the statistical error of flux in that part. So, we do not
consider the fit uncertainties any further.

The deviations of the data from the fit should ideally be Poissonian.
However, we cannot exclude apriori the existence of internal background
variations on any time scale. So, we want to place an upper limit on such
internal variations. We proceed as follows.

a) We choose the time scale 300~s for this study because this is the width
of time bins near 1000~s in Fig.~2 and~3.

b) We average the observed flux in the 300~s bins and make a histogram of
deviations from the fit expressed in units of Poisson error in this bin. In
the absence of internal variations, this histogram should be consistent with
the Gaussian with zero mean and standard deviation = 1 (i.e., with the
normal distribution).

c) We do find that the histogram can be described by normal distribution.
This shows that there are no biases in the background determination and that
the internal background variations on the 300s time scale are small. To set
the upper limit, we fit the width of the distribution and then convert the
upper limit of the width into the corresponding count rate, assuming that
internal variations (if any) and Poisson noise were added in quadrature.

This technique results in a 95\% limit of 0.6 cnt/s for internal background
variations on the 300~s time scale.

\end{document}